\documentclass[pra,reprint,superscriptaddress,showpacs,amsmath, floatfix,amssymb,aps]{revtex4-1}
\makeatletter

\newcommand{\Rmnum}[1]{\expandafter\@slowromancap\romannumeral #1@}
\makeatother
\usepackage{graphicx}% Include figure files
\usepackage{subfigure}
\usepackage{dcolumn}% Align table columns on decimal point
\usepackage{bm}% bold math
\usepackage{hyperref}% add hypertext capabilities
\usepackage{indentfirst}
\usepackage{braket}
\usepackage{amsmath}
\usepackage{ulem}
\usepackage{color}
\hypersetup{colorlinks=true, citecolor=blue, urlcolor=blue, linkcolor=blue}
\begin{document}
\title{Generalized effective-potential Landau theory for a tunable state-dependent hexagonal optical lattice}
\author{Sheng Yue}\thanks{These authors contributed equally to this work.}
\affiliation{School of Physics and optoelectronics engineering, Anhui University, Hefei 230601, P. R. China}
\author{Dan-Yang Chen} \thanks{These authors contributed equally to this work.}
\affiliation{Basic experimental teaching center, Anhui SanLian University, Hefei 230601, P. R. China}
\author{Chenrong Liu}
\affiliation{College of Mathematics and Physics, Wenzhou University, Zhejiang 325035, P. R. China}
\author{Ming Yang}
\affiliation{School of Physics and optoelectronics engineering, Anhui University, Hefei 230601, P. R. China}
\author{Zhi Lin}
\email{zhilin18@ahu.edu.cn}
\affiliation{School of Physics and optoelectronics engineering, Anhui University, Hefei 230601, P. R. China}
\affiliation{State Key Laboratory of Surface Physics and Department of Physics, Fudan University, Shanghai 200433, P. R. China}

\begin{abstract}
We analytically study the ground-state phase diagrams of ultracold bosons with various values of the effective magnetic quantum number $m$ in a state-dependent hexagonal optical lattice  by using the generalized effective-potential Landau theory,  where the site-offset energy between the two triangular sublattice A and B is tunable. Our analytical calculations of third-order corrections  are in reasonably good agreement with the previous cluster Gutzwiller calculations. Furthermore, we reveal the reason why the regions of the Mott lobes $(n,n)$ in phase diagrams for $m=0.02$  are unexpectedly expanded with increasing $J/U$ in deep lattice.
\end{abstract}
\maketitle
\section{Introduction}
Graphene has a profound impact in condensed matter physics owing its linear band dispersion around the Dirac point \cite{graphene1,graphene2}. In solids, many properties of the electrons,  such as the effective mass, the velocity and their interactions, are rigidly determined by the sample itself \cite{Creating-D}, and it is hardly  changed. Consequently, it restricts the potential application of  graphene. In recent years, stimulated by the extraordinary properties of graphene, the emulation of graphene-like physics  in artificial hexagonal or hexagonal-like lattices systems (artificial graphene) is booming \cite{Artificial-graph}.  In progress of the engineering novel optical lattices \cite{engineering-OL} ,  ultracold atoms in hexagonal-like optical lattices provide an avenue to exploring the novel phenomenon which hardly exists in graphene \cite{Creating-D,hexagonal1,hexagonal2,hexagonal3,hexagonal4,hexagonal5,hexagonal6,hexagonal7,hexagonal8}.

In ultracold systems, bosons in the state-dependent hexagonal optical lattice  is the first realization of  an artificial graphene system \cite{hexagonal1}. Here the site-offset energy between two sublattice is dependent on the lattice depth and the value of  the effective magnetic quantum number $m$ of bosons, moreover its strength can be easily tuned. In this novel optical lattice, the phase diagrams  of the single-component quantum gas with various values of $m$ have been numerically studied using cluster Gutzwiller method \cite{hexagonal-Gutzwiller}. Furthermore, the numerical results indicate that the phase diagrams strongly depend on the values of $m$.  However, there is still a lack of analytical results for the phase diagrams of Bose gas in state-dependent hexagonal optical lattice.

So far, the phase diagrams of ultracold bosons in complex optical lattice (such as hexagonal, Kagom\'e optical lattice,  and optical superlattice induced by dipolar interaction)  could have been obtained by several analytical methods, including mean-field theory \cite{SC1}, random-phase approximation \cite{RPA}, strong-coupling expansion (SCE) method \cite{SC1,SC2,SC3}, and the generalized Green's function method  \cite{GGF1,GGF2}, the generalized effective-potential Landau theory (GEPLT) \cite{wang,zhi-4,zhi-5}.  Although the effect of high-order corrections is more obvious in a system with small coordination number, there are still only two analytical methods, i.e., SCE method and GEPLT, which are easy to calculate the high-order corrections of the phase boundaries.

There are some technical issues \cite{zhi-4} in the former GEPLT \cite{wang,wei}, but fortunately these issues are clarified in our previous work \cite{zhi-4}. We have revealed that the clarified GEPLT is extremely well for obtaining the phase diagrams of ultracold bosons in two different type  bipartite superlattices \cite{zhi-4,zhi-5}: one is a bipartite superlattice caused by superlattice structure, and the other one is a bipartite superlattice induced by dipolar interaction (induced bipartite superlattice). More specifically, the phase boundaries of the third-order corrections obtained by the clarified GEPLT are in excellent agreement with the quantum Monte Carlo (QMC) simulations in bipartite superlattice \cite{zhi-4}. In induced bipartite superlattice, the third-order results obtained by the clarified GEPLT are better than the third-order SCE calculations for weak nearest-neighbor repulsion \cite{zhi-4} and these results are also in complete agreement with QMC calculations for second-order phase transition \cite{zhi-5}.

In this paper, with the help of GEPLT, we can study the phase diagrams of single-component quantum gas with  different $m$, i.e., $m=0$, $m=0.02$ and $m=0.1$, in state-dependent hexagonal optical lattice.  In our previous work \cite{zhi-4}, the phase boundaries of the bosons in the normal bipartite superlattice, in which the site-offset energy $\epsilon$ is independent on the trapping lattice depth, have been analytically calculated by the GEPLT. Compared with QMC simulations of this system, we find that our third-order analytical calculations are in excellent agreement with QMC simulations \cite{zhi-4}. Therefore, we are sure that the GEPLT will be a high-accuracy analytical method  for studying the phase diagrams of ultracold bosons in a state-dependent hexagonal optical lattice, in which the site-offset energy $\epsilon$ is dependent on lattice depth and $m$.

\section{the model and method}
The state-dependent hexagonal optical lattice has been realized in the experiment \cite{hexagonal1}, and the total trapping potential \cite{hexagonal1,hexagonal-Gutzwiller,bbm-Gutzwiller} reads
\begin{equation}
V(\textbf{r})=-V_{0}\left[V_{\rm{Hex}}+V_{\rm{pol}}\right],
\end{equation}
where $V_{0}$ is the corresponding lattice depth. Here the state-independent potential reads
\begin{equation}
V_{\rm{Hex}}=6-2\sum_{i}\cos(\mathbf{b}_{i}\mathbf{r}),
\end{equation}
and state-dependent potential for $|F,m_{F}\rangle$ state  reads
\begin{equation}
V_{\rm{Hex}}=\sqrt{3}(-1)^F m_{F} cos\alpha \, \eta\sum_{i}\sin(\mathbf{b}_{i}\mathbf{r}),
\end{equation}
with $\mathbf{b}_{1}=2\pi/\lambda(\sqrt{3},0,0)$, $\mathbf{b}_{2}=\pi/\lambda(-\sqrt{3},3,0)$ and $\mathbf{b}_{3}=\pi/\lambda(-\sqrt{3},-3,0)$, where $\alpha$  is the angle between the quantization axis (or orientation of homogeneous magnetic field $\mathbf{B}$) and lattice plane, and $\eta=0.13$ is dimensionless proportionality factor \cite{hexagonal-Gutzwiller}. Here we can define a effective magnetic quantum number $m=(-1)^{F+1}m_{F} cos\alpha$ \cite{hexagonal-Gutzwiller}, and it is clear that the ranging of $m$ can continuously be adjusted from $-m_{F}$ to $m_{F}$. In Ref.~\cite{hexagonal1}, the wavelength of the lattice laser beams is chosen as $\lambda=830$ nm, and the corresponding lattice constant is $a=2\lambda/(3\sqrt3)$.

In  Dirk-S\"oren L\"uhmann et al.'s work \cite{hexagonal-Gutzwiller}, the optimal Wannier functions are determined in such a way that the amplitudes
of processes beyond the Hubbard model are minimized. By choosing this optimal Wannier functions, the corresponding Hamiltonian of the state-dependent hexagonal optical lattice can read \cite{hexagonal-Gutzwiller}
\begin{equation}
\hat{H}=-J\!\sum_{\langle i,j\rangle}\!\hat{a}_{i}^{\dagger}\hat{a}_{j}+\!\frac{1}{2}\sum_{i}\!U_{i}\hat{n}_{i}(\hat{n}_{i}-1)+
\sum_{i}(\epsilon_{i}-\mu)\hat{n}_{i}, \label{SBH}
\end{equation}
where $J$ is the nearest-neighbor tunneling, $\hat{a}_{i}$ ($\hat{a}_{i}^{\dagger}$) is the boson  annihilation(creation) operator at site $i$, $\hat{n}_{i}= \hat{a}_{i}^{\dagger}\hat{a}_{i}$ is the particle-number operator on $i$-th site, $U_{i}$ denotes the on-site repulsion at site $i$, $\mu$ is the chemical potential, and $\epsilon_{i}$ is the site-offset energy($\epsilon_{A}=0$ and $\epsilon_{B}=\epsilon$ for sublattice A and B).
Since only the site-offset energy $\epsilon$ is approximately linearly dependent on the parameter $m$ \cite{hexagonal-Gutzwiller}, but the other interactions are weak dependent on  $m$, i.e., $J(m)\approx J$ and $U_{A}(m)\approx U_{B}(m)\approx U$ (see the Figs. 8 in the ref.~\cite{hexagonal-Gutzwiller}), and therefore the phase boundaries in the ref.~\cite{hexagonal-Gutzwiller} can be obtained by using cluster Gutzwiller method to study the standard Hubbard model, in which the on-site repulsions $U_{A}$ and $U_{B}$ are assumed to be equal.

\begin{figure}[h!]
\centering
\includegraphics[width=1.0\linewidth]{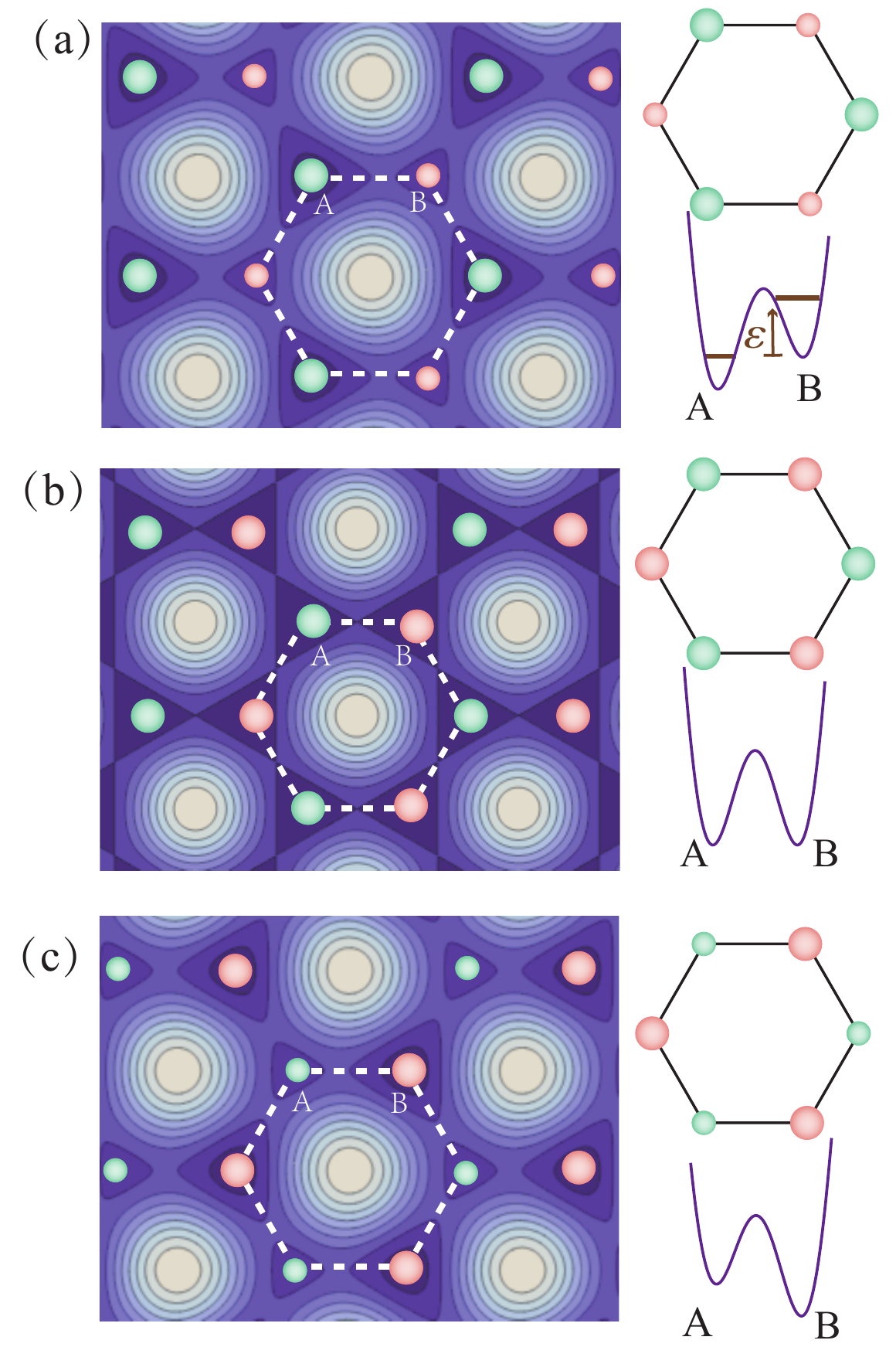}
\caption{The sketch of the state-dependent hexagonal optical lattice with different effective magnetic quantum number $m$, where $m$ are equal to 1, 0 and -1 for (a), (b) and (c). The site-offset energy  $\varepsilon$ between the two triangular sublattice A and B  is is tunable.}
\label{1}
\end{figure}

\begin{figure}[h!]
\centering
\includegraphics[width=0.9\linewidth]{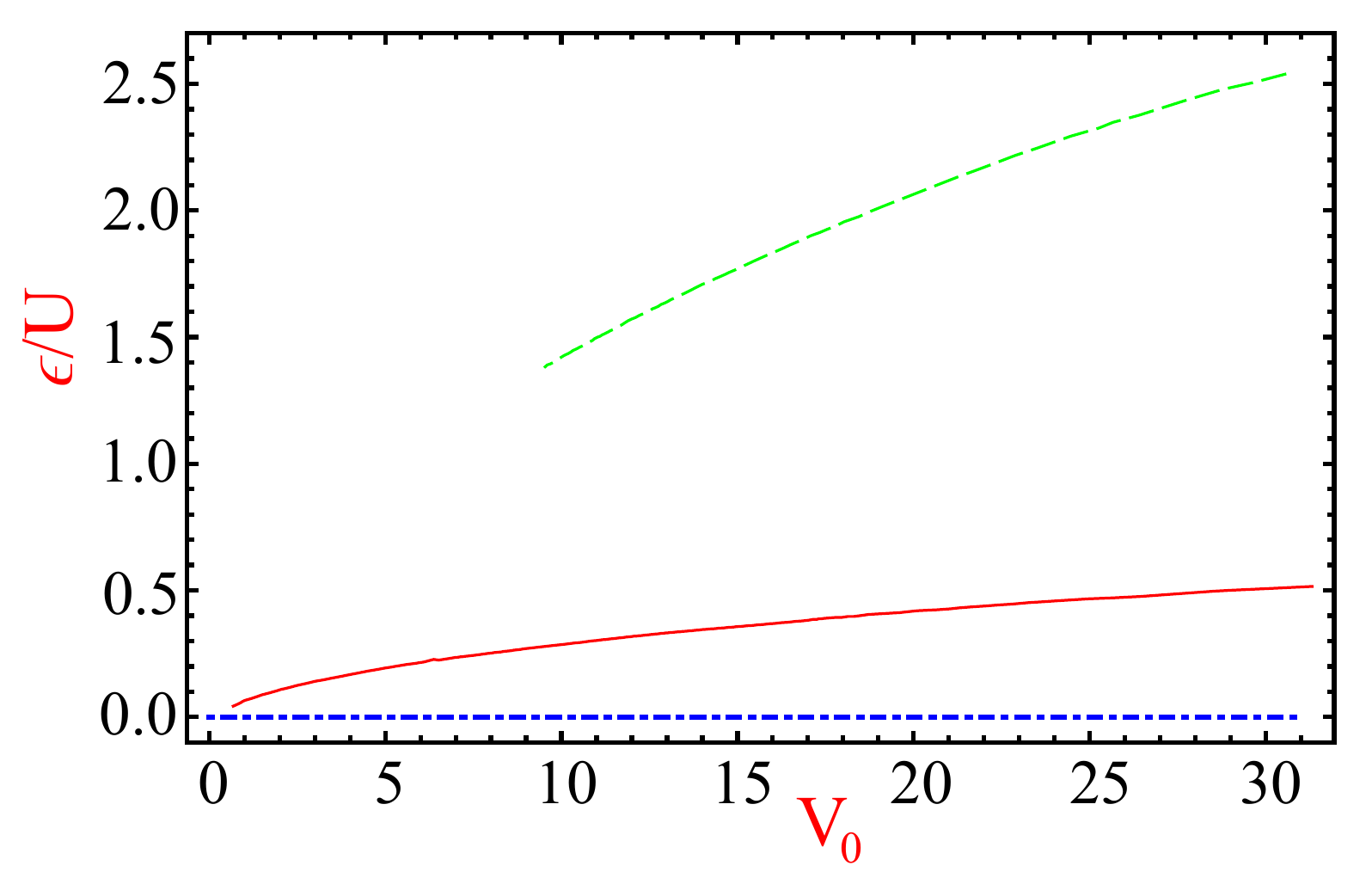}
\caption{ The site-offset energies $\epsilon/U$ with different effective magnetic quantum number $m$ as function of the lattice depth $V_{0}$, where the red line,  the green dashed line  and the blue dot-dashed line  represents  $\epsilon(m=0.1)/U$, $\epsilon(m=0.02)/U$ and $\epsilon(m=0)/U$( sublattice A and B are balanceable), respectively. The data for $\epsilon(m=0.1)/U$ and  $\epsilon(m=0.02)/U$ come from the ref.~\cite{hexagonal-Gutzwiller}. Here we do not draw the curve of $\epsilon(m=0.1)/U$ within the regions of $V_{0}\lesssim 9.5$, owing to the fact that these data are not provided in ref.~\cite{hexagonal-Gutzwiller}.}
\label{2}
\end{figure}

Before calculating the phase diagrams, we would like to discuss the possible ground states of the systems in the atomic limit ($J=0$). In this limit, the ground states are determined by the competing between the on-site interaction $U$ and  the site-offset energy $\epsilon$. If the Mott lobe $(n_A,n_B)$ with the imbalance $\Delta n=n_{A}-n_{B}$ and the fixed filling $\rho=(n_{A}+n_{B})/2$ is in existence, $\epsilon$ needs to meet the constraint $\Delta n-1<\epsilon/U<\Delta n+1$ obtained by solving the inequations $E_{n_{A}+1,n_{B}-1}>E_{n_{A},n_{B}} <E_{n_{A}-1,n_{B}+1}$, where $n_{B}$ is nonzero. In the case of $n_{B}=0$,  the Mott lobe $(1,0)$, i.e., $\Delta n=1$,  can always exist in  the systems with $\epsilon/U>0$.   The site-offset energies $\epsilon(m=0)/U$, $\epsilon(m=0.02)/U$ and $\epsilon(m=0.1)/U$ as a function of the variational trapping lattice depth $V_{0}$ are shown in Fig.~\ref{2}, respectively.  In the deep lattice depth, due to the $\epsilon(m=0.02)/U$ in the range of $(0,1)$, therefore the $\Delta n$ for the Mott lobes $(n_A,n_B)$ can take two values $\Delta n=0$ and  $\Delta n=1$. For the similar reason,  the $\Delta n$ can take three values, i.e., $\Delta n=1$, $\Delta n=2$, and $\Delta n=3$ in the system with $m=0.1$, where we have $2<\epsilon(m=0.1)/U<3$. Taking $m=0.02$ as an example, we will analyse the chemical potential width of the Mott lobes $(n_{A},n_{B})$. In the Mott lobes $(n,n)$ [$(n,n-1)$], the chemical potential should satisfy constraint inequations $E_{n,n-1}\ge E_{n,n} \le E_{n+1,n}$ ($E_{n-1,n-1}\ge E_{n,n-1} \le E_{n,n}$), i.e., $\epsilon +U(n-1)\le \mu \le Un$ [$U(n-1)\le \mu \le \epsilon +U(n-1)$], where the $E_{n,n}$ is the eigenvale of the $\hat{H}(J=0)$. More specifically, the chemical potential widths of the Mott lobes $(1,0)$, $(1,1)$, $(2,1)$, $(2,2)$, $(3,2)$ and $(3,3)$ are $0\le \mu \le \epsilon$, $\epsilon\le \mu \le U$, $U\le \mu \le \epsilon+U$,  $\epsilon+U\le \mu \le 2U $, $2U \le \mu \le \epsilon+2U $, and $\epsilon+2U  \le \mu \le 3U $, respectively.

We can make a reasonable assumption that the features of the Mott lobes obtained in the limit $J=0$ will not change when we keep $J/U\ll 1$.
Hence, the region of Mott lobes $(n,n)$ will expand with increasing  the value of $J/U$ (or decreasing the value of the lattice depth) in the deep lattice ($J/U\ll 1$), owing the fact that $\epsilon$ is decreasing with increasing the value of $J/U$. This feature of  Mott lobes $(n,n)$ is in contrast to the feature of  Mott lobes $(n,n)$ in normal superlattice system in which the $\epsilon$ is constant (do not change with the lattice depth), and this feature has clearly been shown in Fig.~ \ref{phase-1}(b) [or see the Fig. 9(b) in ref.~\cite{hexagonal-Gutzwiller}].

\begin{figure}[h!]
\centering
\includegraphics[width=1.0\linewidth]{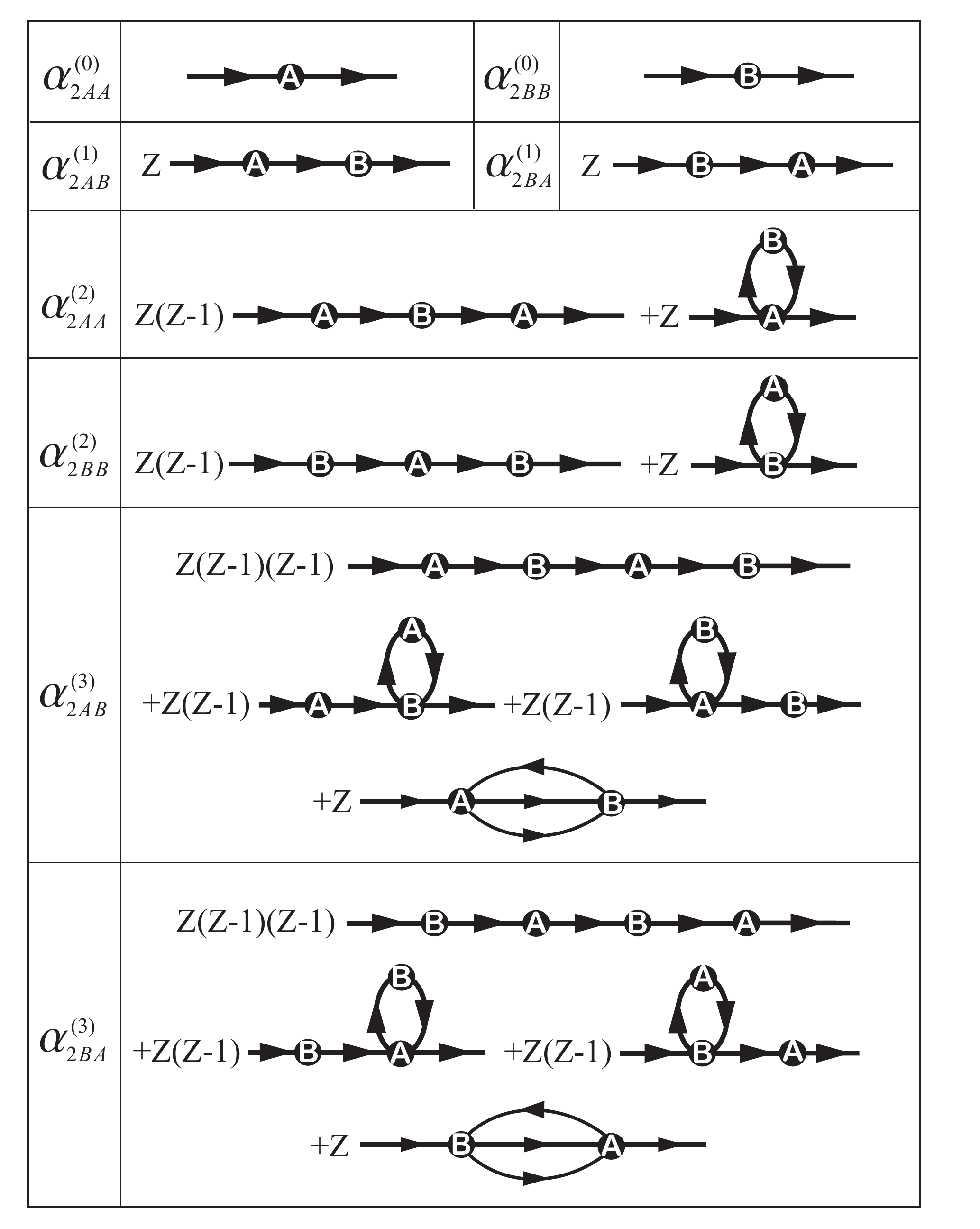}
\caption{The diagrams of the coefficients $\alpha^{(n)}_{2ij}$ for a tunable state-dependent hexagonal optical lattice, where $Z=3$ is the coordination number of the systems. The diagrams of the coefficients $\alpha^{(n)}_{2ij}$ in state-dependent hexagonal lattice are the same with these figures in normal bipartite superlattices \cite{zhi-4,zhi-5}.}
\label{table-1}
\end{figure}

As mentioned above,  the features of the Mott lobes have been revealed  in the atomic limit ($J=0$). At below, we will introduce the GEPLT to study the phase boundaries of single-component quantum gas in state-dependent hexagonal optical lattice, in which the site-offset energy $\epsilon$ is dependent on lattice depth $V_0$ and effective magnetic quantum number $m$.  In our previous work \cite{zhi-4}, we have revealed that the GEPLT is good enough to obtain phase boundaries of bipartite superlattice with the fixed site-offset energy $\Delta \mu $ which is independent on $V_0$. By mapping $\mu\rightarrow \mu +\epsilon$ and $\epsilon \rightarrow \Delta \mu$, then the forms of the Hamiltonian in state-dependent hexagonal optical lattice and the normal bipartite superlattice are the same. Thus, we can directly use the GEPLT, which is established in normal bipartite superlattice system with the fixed site-offset energy $\Delta \mu $, to study the phase diagrams of the state-dependent hexagonal systems. Immediately, we can obtain the first three orders of the phase boundaries $J^{(n)}_{c}$ of single-component quantum gas in state-dependent hexagonal optical lattice, which read
\begin{equation}
J^{(1)}_{c}=\sqrt{\alpha_{2AA}^{(0)}\alpha_{2BB}^{(0)}}/\alpha_{2AB}^{(1)},\label{mean-field}
\end{equation}
\begin{equation}
J^{(2)}_{c}=\frac{2\sqrt{\alpha_{2AA}^{(0)}\alpha_{2BB}^{(0)}}\alpha_{2AB}^{(1)}}{\alpha_{2AA}^{(0)}\alpha_{2BB}^{(2)}+\alpha_{2BB}^{(0)}\alpha_{2AA}^{(2)}},
\label{2-order}
\end{equation}
\begin{equation}
J^{(3)}_{c}=\frac{\alpha_{2AA}^{(0)}\alpha_{2BB}^{(2)}+\alpha_{2BB}^{(0)}\alpha_{2AA}^{(2)}}{2\sqrt{\alpha_{2AA}^{(0)}\alpha_{2BB}^{(0)}}\alpha_{2AB}^{(3)}},
\label{3-order}
\end{equation}
where $\alpha^{(n)}_{2ij}$ are perturbative coefficients for state-dependent hexagonal systems. With the help of  Raylieigh-Schr\"{o}dinger perturbation expansion, we can easily calculate the values of  these perturbative coefficients $\alpha^{(n)}_{2ij}$. The specific process of calculating the perturbative coefficients $\alpha^{(n)}_{2ij}$ have  been clearly demonstrated \cite{santos,zhi-1,wang,zhi-5}.

\section{The ground-state phase diagram}
\begin{figure}[h!]
\centering
\includegraphics[width=1.0\linewidth]{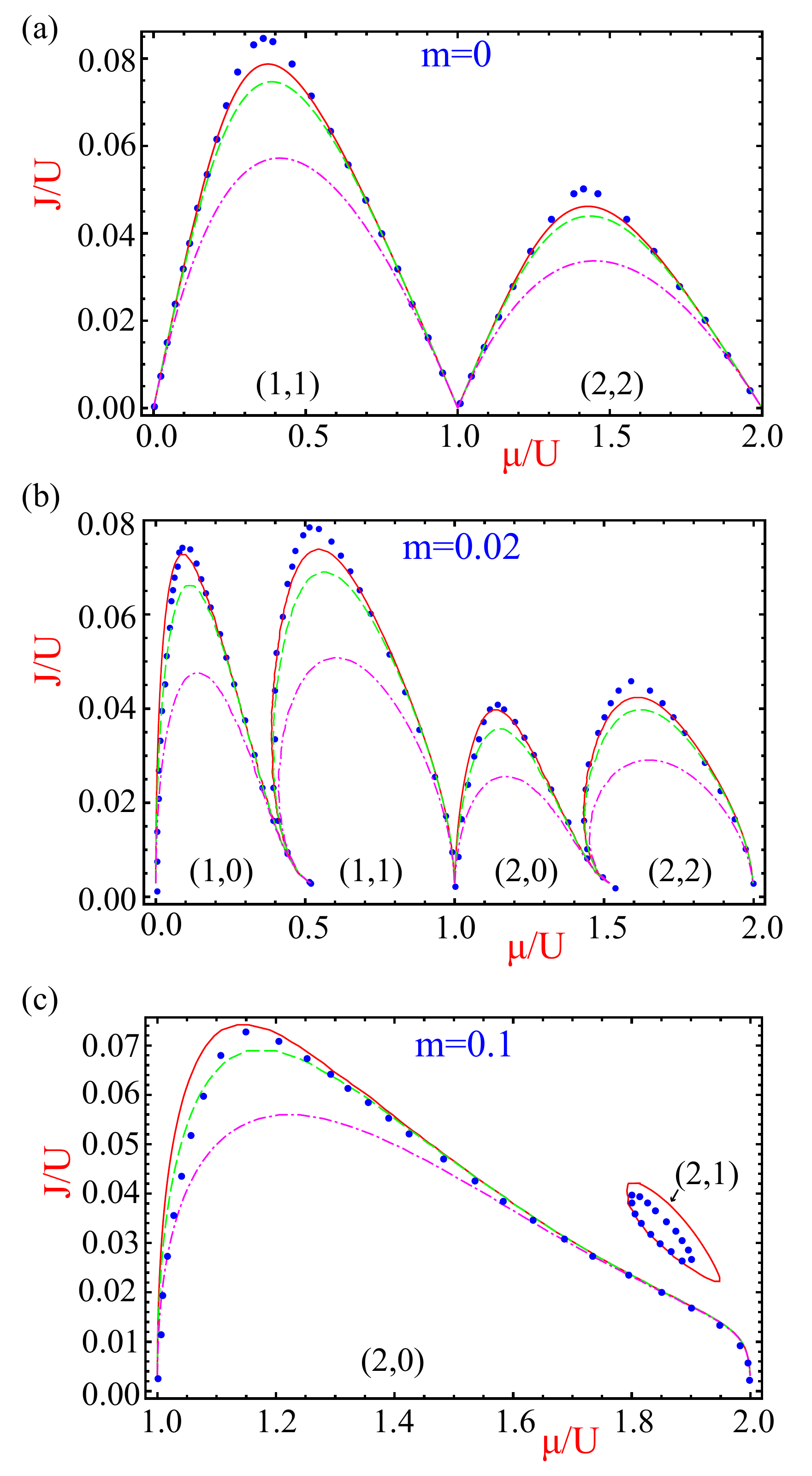}
\caption{The phase diagrams of single-component quantum gas in state-dependent hexagonal optical lattice for  effective magnetic quantum number $m=0$(a)\cite{zhi-1}, $m=0.02$(b) and $m=0.1$(c). The red solid lines are the third-order results, the green dashed lines are the second-order calculations, and the pink dot-dashed line are the first-order (mean-field) estimations. Here the blue dots are results obtained by cluster Gutzwiller calculation \cite{hexagonal-Gutzwiller}.}
\label{phase-1}
\end{figure}
With the help of the phase boundaries in Eqs.~(\ref{mean-field})-(\ref{3-order}), we can analytically obtain the first three orders of the phase boundaries with effective magnetic quantum  number $m=0$, $m=0.02$ and $m=0.1$, respectively.

\begin{figure}[h!]
\centering
\includegraphics[width=1.0\linewidth]{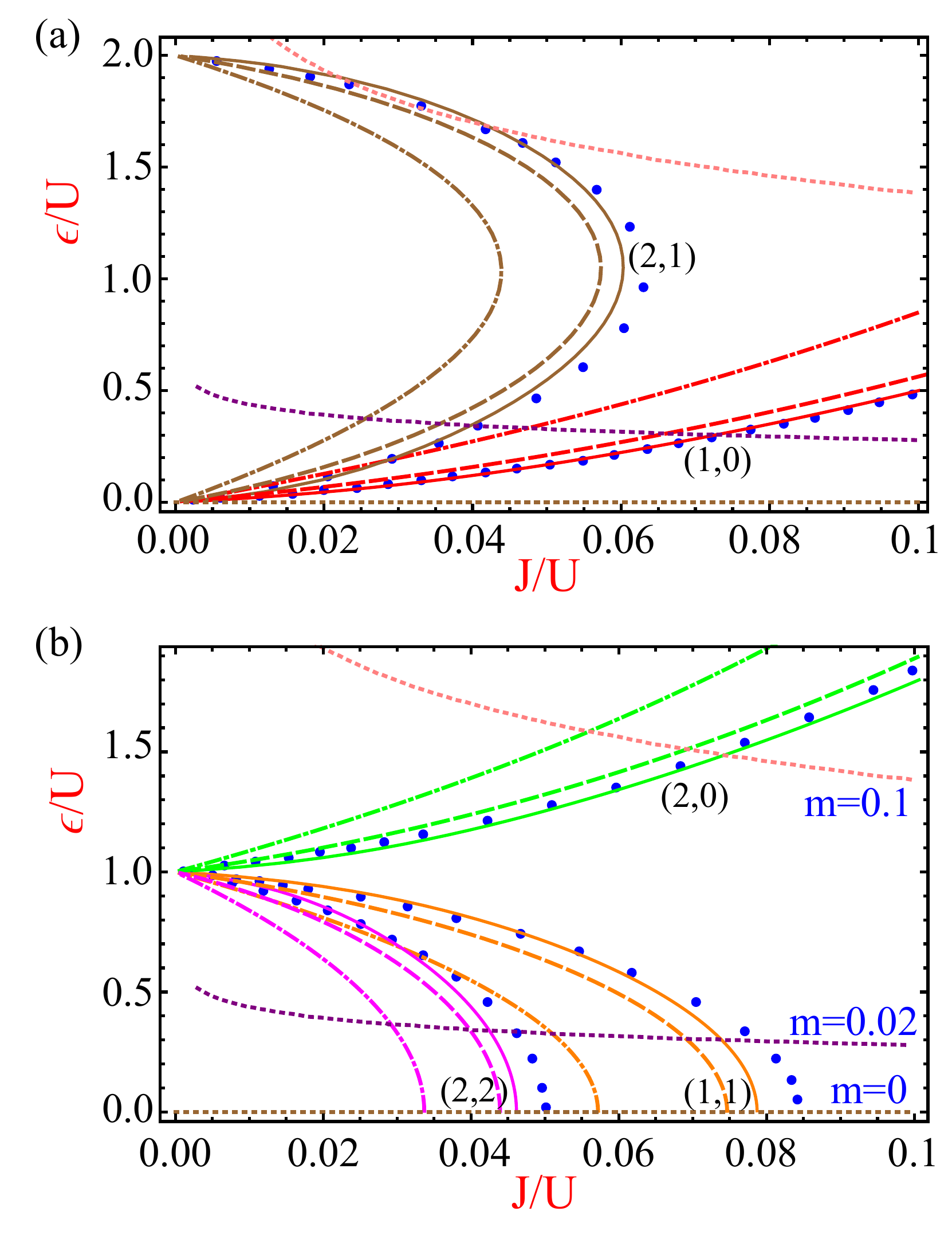}
\caption{The universal $\epsilon/U$-$J/U$ phase diagrams of single-component quantum gas with arbitrary values of the  effective magnetic quantum number $m$ for (a) half-integer and (b) integer filling. Here the blue dots are results obtained by cluster Gutzwiller calculation \cite{hexagonal-Gutzwiller}, and the all solid lines are the third-order results, the all  long-dash lines are the second-order calculations, and the all  dot-dashed lines are the first-order estimation. The different colours denote the different Mott phases, i.e., the brown represents the Mott phase $(2,1)$,  the red represents the Mott phase $(1,0)$,  the green describes the Mott phase $(2,0)$, the magenta describes the Mott phase $(2,2)$ and  the orange describes the Mott phase $(1,1)$. The pink, purple and brown short-dash line denotes $m=0.1$, $m=0.02$, and $m=0$, respectively.}
\label{phase-2}
\end{figure}
From the Fig.~\ref{phase-1},  it is easy to find that our analytical results are consistent with cluster Gutzwiller simulations for effective magnetic quantum number $m=0$ and $m=0.02$. In the case of $m=0.1$, our analytical results are in good accord with  cluster Gutzwiller calculations in the Mott lobe $(2,0)$ for the large interval, and only in small interval $\mu/U\in(1.04,1.15)$, there are some deviations between our analytical results and  cluster Gutzwiller simulations \cite{hexagonal-Gutzwiller}. Moreover, it is interesting that the Mott phase $(2,1)$  forms a island in contrast to  a normal lobe in the parameter space. In order to obtain the outline of the Mott island, we need to consider at least the contributions of third-order corrections, the first two orders of the calculations do not have ability to obtain this interesting Mott island [See Fig.~\ref{phase-1}(c)]. One of the possible reasons why there are some deviations between our results and cluster Gutzwiller calculations for the Mott island $(2,1)$ and Mott lobe $(2,0)$ in small interval $\mu/U\in(1.04,1.15)$ is that the type of phase transitions from Mott island or lobe  to superfluid is dependent on $\mu$. This interesting phenomenon has been revealed in the bipartite lattice system caused by dipolar interaction \cite{Ohgoe}. More specifically, the phase transition is  maybe a fist-order or weak fist-order phase transition for Mott lobe $(2,0)$ [Mott island $(2,1)$] with small (large) $\mu$, but  the phase transition is second-order phase transition in the other regions.

In order to study the deviations of the tips of Mott phases obtained by these two methods, we also reveal the universal $\epsilon/U$-$J/U$ phase diagrams, in which each point $(\epsilon/U,J/U)$ represents the tip of the Mott phase in $J-\mu-U$ phase diagrams for the fixed $\epsilon/U$. From the Fig.~\ref{phase-2}, it is clear that the values of the tips obtained by our analytical method   are perfectly consistent with  cluster Gutzwiller simulations, the relative deviations of our third-order results and cluster Gutzwiller calculations are less than $9\%$($5\%$) for Mott insulator with (half-)integer filling.

\section{Conclusion}
The phase diagrams of single-component Bose gas with various values of $m$ in the state-dependent hexagonal optical have been revealed by using the GEPLT. In the case of $m=0.02$, we reveal that the Mott lobes $(n,n)$ will abnormally expand with increasing $J/U$ in deep lattice.  The reason why there exists such abnormally expanding phenomenon is that the site-offset energy $\epsilon$ decreases with increasing $J/U$. Furthermore, this interesting  phenomenon can be detected in this novel lattice with a harmonic confinement by using in situ imaging techniques \cite{Ott}.  In the case of $m=0.1$, we find that our third-order analytical results can well demonstrate that the outline form of the Mott phase $(2,1)$ is island.  More generally, the first three orders of the corrections have also been calculated analytically, and our third-order analytical results are in good agreement with the numerical solutions obtained by cluster Gutzwiller method. Besides, the values $J_c/U$ of the tips in the Mott phases as a function of the site-offset energy $\epsilon/U$ have also been  calculated analytically. Compared to cluster Gutzwiller simulations, the relative deviations of these tips of our third-order results  are less than $9\%$ ($5\%$) for Mott insulator with (half-)integer filling. Thus, the GEPLT as an accurate analytical method, which is easy to obtain higher orders hopping corrections, has tremendous potential especially for studying the second-order quantum phase transition of ultracold bosons in lattices with small coordination numbers, for example, artificial hexagonal-like lattices systems.

\section*{Acknowledgement}
This research is supported by the National Natural Science Foundation of China (NSFC) under Grant Nos.11947102 and 12004005, the Natural Science Foundation of Anhui Province under Grant Nos. 2008085QA26 and 2008085MA16, and the Ph.D. research Startup Foundation of Wenzhou University under Grant No. KZ214001P05, and the open project of state key laboratory of surface physics in Fudan University (Grant No. KF2021$\_$08).


\begin{thebibliography}{99}
\bibitem{graphene1}A. H. Castro Neto, F. Guinea, N. M. R. Peres,  K. S. Novoselov, and A. K. Geim,  The electronic properties of graphene, Rev. Mod. Phys. \textbf{81}, 109 (2009).
\bibitem{graphene2}M. Z. Hasan, and C. L. Kane, Topological insulators, Rev. Mod. Phys. \textbf{81}, 109 (2009).
\bibitem{Creating-D}L. Tarruell, D. Greif, T. Uehlinger, G. Jotzu, and T. Esslinger, Creating, moving and merging Dirac points with a
Fermi gas in a tunable honeycomb lattice, Nature (London) \textbf {483}, 302 (2012).
\bibitem{Artificial-graph} M. Polini, F. Guinea, M. Lewenstein, H. C. Manoharan, and V. Pellegrini, Artificial honeycomb lattices for electrons, atoms and photons, Nat. Nanotechnol. \textbf{8}, 625 (2013).
\bibitem{engineering-OL} P. Windpassinger, and K. Sengstock, Engineering novel optical lattices, Rep. Prog. Phys. \textbf{76}, 086401 (2013).
\bibitem{hexagonal1}P. Soltan-Panahi, J. Struck, P. Hauke, A. Bick, W. Plenkers, G. Meineke, C. Becker,
P.Windpassinger, M. Lewenstein, and K. Sengstock, Multi-component quantum gases in
spin-dependent hexagonal lattices, Nat. Phys. \textbf{7}, 434 (2011).
\bibitem{hexagonal2}P. Soltan-Panahi, D.-S. L\"uhmann, J. Struck, P. Windpassinger, and K. Sengstock, Quantum phase transition to unconventional multi-orbital superfluidity in optical lattices, Nat. Phys. \textbf{8}, 71 (2012).
\bibitem{hexagonal3}T. Uehlinger, G. Jotzu, M. Messer, D. Greif, W. Hofstetter, U. Bissbort, and T. Esslinger, Artificial Graphene with Tunable Interactions, Phys. Rev. Lett. \textbf{111}, 185307 (2013).
\bibitem{hexagonal4} G. Jotzu, M. Messer, R. Desbuquois, M. Lebrat, T. Uehlinger, D. Greif, and T. Esslinger, Experimental realization of the topological Haldane model with ultracold fermions, Nature (London) \textbf {515}, 237 (2014).
\bibitem{hexagonal5} L. Duca, T. Li, M. Reitter, I. Bloch, M. Schleier-Smith, and U. Schneider, An Aharonov-Bohm interferometerfor determining Bloch band topology,  Science \textbf{347}, 288 (2015).
\bibitem{hexagonal6} M. Messer, R. Desbuquois, T. Uehlinger, G. Jotzu, S. Huber, D. Greif, and T. Esslinger, Exploring Competing Density Order in the Ionic Hubbard Model with Ultracold Fermions, Phys. Rev. Lett. \textbf{115}, 115303 (2015).
\bibitem{hexagonal7} N. Fl\"aschner, B. S. Rem, M. Tarnowski, D. Vogel, Dirk-S\"oren L\"uhmann, K. Sengstock, and C. Weitenberg, Experimental reconstruction of the Berry curvature in a Floquet Bloch band, Science \textbf{352}, 1091 (2016).
\bibitem{hexagonal8}T. Li, L. Duca, M. Reitter, F. Grusdt, E. Demler, M. Endres, M. Schleier-Smith, I. Bloch, and U. Schneider, Bloch state tomography using Wilson lines, Science \textbf{352}, 1094 (2016).
\bibitem{hexagonal-Gutzwiller} Dirk-S\"oren L\"uhmann, O. J\"urgensen, M. Weinberg, J. Simonet, P. Soltan-Panahi, and K. Sengstock, Phys. Rev. A \textbf{90}, 013614 (2014).
\bibitem{SC1}M. Iskin, and J. K. Freericks, Strong-coupling perturbation theory for the extended Bose-Hubbard model, Phys. Rev. A \textbf{79}, 053634 (2009).
\bibitem{SC2} M. Iskin, Strong-coupling expansion for the two-species Bose-Hubbard model, Phys. Rev. A \textbf{82}, 033630(2010).
\bibitem{SC3}O. J\"urgensen, and Dirk-S\"oren L\"uhmann, Dimerized Mott insulators in hexagonal optical lattices, New J. Phys. \textbf{16},  093023 (2014).
\bibitem{RPA} M. Iskin, and J. K. Freericks, Momentum distribution of the insulating phases of the extended Bose-Hubbard model, Phys. Rev. A \textbf{80}, 063610 (2009).
\bibitem{GGF1}J. Zhang and Y. Jiang, Quantum phase diagrams and time-of-flight pictures of spin-1 Bose systems in honeycomb optical lattices, Laser Phys. \textbf{26}, 095501 (2016).
\bibitem{GGF2}Z. Lin, J. Zhang, and Y. Jiang, Analytical approach to quantum phase transitions of ultracold Bose gases in bipartite optical lattices using the generalized Green's function method, Front. Phys. \textbf{13}(4), 136401 (2018).
\bibitem{wang} T. Wang, X. F. Zhang, S. Eggert, and A. Pelster, Generalized effective-potential Landau theory for bosonic quadratic superlattices, Phys. Rev. A \textbf{87}, 063615 (2013).
\bibitem{zhi-4} Z. Lin, and W. L. Liu, Analytic calculation of high-order corrections to quantum phase transitions of ultracold Bose gases in bipartite superlattices, Front. Phys. \textbf{13}(5), 136402 (2018).
\bibitem{zhi-5} Z. Lin, and M. Yang, Generalized effective-potential Landau theory for the two-dimensional extended Bose-Hubbard model, Phys. Lett. A \textbf{383}, 1666 (2019).
\bibitem{wei}F. Wei, J. Zhang, and Y. Jiang, Quantum phase diagram and time-of-flight absorption pictures of an ultracold
Bose system in a square optical superlattice, Europhys. Lett. \textbf{113}, 16004 (2016).
%\bibitem{QMC11}T. Ohgoe, T. Suzuki, and N. Kawashima, Phys. Rev. B \textbf{86}, 054520 (2012).
\bibitem{bbm-Gutzwiller}L. S. Cao, S. Kr\"onke, J. Stockhofe, J. Simonet, K. Sengstock, Dirk-S\"oren L\"uhmann, and P. Schmelcher, Beyond-mean-field study of a binary bosonic mixture in a state-dependent honeycomb lattice,  Phys. Rev. A \textbf{91}, 043639 (2015).
\bibitem{santos}F. E. A. dos Santos, and A. Pelster,  Quantum phase diagram of bosons in optical lattices, Phys. Rev. A \textbf{79}, 013614 (2009).
\bibitem{zhi-1} Z. Lin, J. Zhang, and Y. Jiang, Quantum phase transitions of ultracold Bose systems in nonrectangular optical lattices,  Phys. Rev. A \textbf{85}, 023619 (2012).
\bibitem{Ohgoe}T. Ohgoe, T. Suzuki, N. Kawashima, Ground-state phase diagram of the two-dimensional extended Bose-Hubbard model, Phys. Rev. B \textbf{86}, 054520 (2012).
\bibitem {Ott}H. Ott, Single atom detection in ultracold quantum gases: a review of current progress, Rep. Prog. Phys. \textbf{79}, 054401 (2016).
\end{thebibliography}
\end{document}